\input hyperbasics
\catcode`\@=11
\def\unredoffs{\voffset=15truemm \hoffset=10truemm} 
\def\redoffs{\voffset=-12.truemm\hoffset=-9truemm} 
\def\speclscape{}
\def\redoffs{\voffset=-12.truemm\hoffset=-14truemm} 
\newbox\leftpage \newdimen\fullhsize \newdimen\hstitle \newdimen\hsbody
\newdimen\hdim
\hfuzz=1pt
\ifx\hyperdef\UNd@FiNeD\def\hyperdef#1#2#3#4{#4}\def\hyperref#1#2#3#4{#4}\fi
\def\newans{y }
\def\answb{y }
\ifx\answb\newans\message{(This uses normal fonts.)}%
\magnification=1200
%
\def\bigans{b }
\def\answ{b }
\ifx\answ\bigans\message{(Format simple colonne 12pts.}
\magnification=1200 
\unredoffs\hsize=120mm\vsize=6.68in
\hsbody=\hsize \hstitle=\hsize 
\else\message{(Format double colonne, 10pts.} \let\l@r=L
\magnification=1000 \vsize=6.68in
\redoffs%
\hstitle=120truemm\hsbody=120truemm\fullhsize=261.5truemm\hsize=\hsbody 
\output={
  \almostshipout{\leftline{\vbox{\makeheadline\pagebody\makefootline}}}
\advancepageno%
}
\def\almostshipout#1{\if L\l@r \count1=1 \message{[\the\count0.\the\count1]}
      \global\setbox\leftpage=#1 \global\let\l@r=R
 \else \count1=2
  \shipout\vbox{\speclscape{\hsize\fullhsize}
      \hbox to\fullhsize{\box\leftpage\hfil#1}}  \global\let\l@r=L\fi}
\fi

\def\sla#1{\mkern-1.5mu\raise0.4pt\hbox{$\not$}\mkern1.2mu #1\mkern 0.7mu}
\def\Dbar{\mkern-1.5mu\raise0.4pt\hbox{$\not$}\mkern-.1mu {\rm D}\mkern.1mu}
\def\Abar{\mkern1.mu\raise0.4pt\hbox{$\not$}\mkern-1.3mu A\mkern.1mu}
\def\Bbar{\mkern-0.mu\raise0.4pt\hbox{$\not$}\mkern-.3mu B\mkern 0.6mu}
\newskip\tableskipamount \tableskipamount=8pt plus 3pt minus 3pt
\def\tableskip{\vskip\tableskipamount}

\newdimen\chapskip

\font\caprm=cmr9
\font\capit=cmti9
\font\capbf=cmbx9
\font\capsl=cmsl9
\font\capmi=cmmi9
\font\capex=cmex9
\font\capsy=cmsy9
\chapskip=17.5mm
\def\makeheadline{\vbox to 0pt{\vskip-22.5pt
\line{\vbox to8.5pt{}\the\headline}\vss}\nointerlineskip}
\font\tenbi=cmmib10 
\font\ninebi=cmmib9
\font\sevenbi=cmmib7 
\font\fivebi=cmmib5
\textfont4=\tenbi
\scriptfont4=\sevenbi
\scriptscriptfont4=\fivebi

\font\sixrm=cmr6
\font\fiverm=cmr5
\font\sixmi=cmmi6
\font\fivemi=cmmi5
\font\sixsy=cmsy6
\font\fivesy=cmsy5
\font\sixbf=cmbx6
\font\fivebf=cmbx5
\skewchar\capmi='177 \skewchar\sixmi='177 \skewchar\fivemi='177
\skewchar\capsy='60 \skewchar\sixsy='60 \skewchar\fivesy='60

\def\elevenpoint{
\textfont0=\caprm \scriptfont0=\sixrm \scriptscriptfont0=\fiverm
\def\rm{\fam0\caprm}
\textfont1=\capmi \scriptfont1=\sixmi \scriptscriptfont1=\fivemi
\textfont2=\capsy \scriptfont2=\sixsy \scriptscriptfont2=\fivesy
\textfont3=\capex \scriptfont3=\capex \scriptscriptfont3=\capex
\textfont\itfam=\capit \def\it{\fam\itfam\capit} 
\textfont\slfam=\capsl  \def\sl{\fam\slfam\capsl} 
\textfont\bffam=\capbf \scriptfont\bffam=\sixbf
\scriptscriptfont\bffam=\fivebf
\def\bf{\fam\bffam\capbf} 
\textfont4=\ninebi \scriptfont4=\sevenbi \scriptscriptfont4=\fivebi
\abovedisplayskip=11pt plus 3pt minus 8pt
\belowdisplayskip=\abovedisplayskip
\smallskipamount=2.7pt plus 1pt minus 1pt
\medskipamount=5.4pt plus 2pt minus 2pt
\bigskipamount=10.8pt plus 3.6pt minus 3.6pt
\normalbaselineskip=11pt
\setbox\strutbox=\hbox{\vrule height7.8pt depth3.2pt width0pt}
\normalbaselines \rm}

%
%

\catcode`\@=11

\font\tenmsa=msam10
\font\sevenmsa=msam7
\font\fivemsa=msam5
\font\tenmsb=msbm10
\font\sevenmsb=msbm7
\font\fivemsb=msbm5
\newfam\msafam
\newfam\msbfam
\textfont\msafam=\tenmsa  \scriptfont\msafam=\sevenmsa
  \scriptscriptfont\msafam=\fivemsa
\textfont\msbfam=\tenmsb  \scriptfont\msbfam=\sevenmsb
  \scriptscriptfont\msbfam=\fivemsb

\def\hexnumber@#1{\ifcase#1 0\or1\or2\or3\or4\or5\or6\or7\or8\or9\or
	A\or B\or C\or D\or E\or F\fi }

\font\teneuf=eufm10
\font\seveneuf=eufm7
\font\fiveeuf=eufm5
\newfam\euffam
\textfont\euffam=\teneuf
\scriptfont\euffam=\seveneuf
\scriptscriptfont\euffam=\fiveeuf
\def\frak{\ifmmode\let\next\frak@\else
 \def\next{\Err@{Use \string\frak\space only in math mode}}\fi\next}
\def\goth{\ifmmode\let\next\frak@\else
 \def\next{\Err@{Use \string\goth\space only in math mode}}\fi\next}
\def\frak@#1{{\frak@@{#1}}}
\def\frak@@#1{\fam\euffam#1}

\edef\msa@{\hexnumber@\msafam}
\edef\msb@{\hexnumber@\msbfam}

\mathchardef\boxdot="2\msa@00
\mathchardef\boxplus="2\msa@01
\mathchardef\boxtimes="2\msa@02
\mathchardef\square="0\msa@03
\mathchardef\blacksquare="0\msa@04
\mathchardef\centerdot="2\msa@05
\mathchardef\lozenge="0\msa@06
\mathchardef\blacklozenge="0\msa@07
\mathchardef\circlearrowright="3\msa@08
\mathchardef\circlearrowleft="3\msa@09
\mathchardef\rightleftharpoons="3\msa@0A
\mathchardef\leftrightharpoons="3\msa@0B
\mathchardef\boxminus="2\msa@0C
\mathchardef\Vdash="3\msa@0D
\mathchardef\Vvdash="3\msa@0E
\mathchardef\vDash="3\msa@0F
\mathchardef\twoheadrightarrow="3\msa@10
\mathchardef\twoheadleftarrow="3\msa@11
\mathchardef\leftleftarrows="3\msa@12
\mathchardef\rightrightarrows="3\msa@13
\mathchardef\upuparrows="3\msa@14
\mathchardef\downdownarrows="3\msa@15
\mathchardef\upharpoonright="3\msa@16

\mathchardef\downharpoonright="3\msa@17
\mathchardef\upharpoonleft="3\msa@18
\mathchardef\downharpoonleft="3\msa@19
\mathchardef\rightarrowtail="3\msa@1A
\mathchardef\leftarrowtail="3\msa@1B
\mathchardef\leftrightarrows="3\msa@1C
\mathchardef\rightleftarrows="3\msa@1D
\mathchardef\Lsh="3\msa@1E
\mathchardef\Rsh="3\msa@1F
\mathchardef\rightsquigarrow="3\msa@20
\mathchardef\leftrightsquigarrow="3\msa@21
\mathchardef\looparrowleft="3\msa@22
\mathchardef\looparrowright="3\msa@23
\mathchardef\circeq="3\msa@24
\mathchardef\succsim="3\msa@25
\mathchardef\gtrsim="3\msa@26
\mathchardef\gtrapprox="3\msa@27
\mathchardef\multimap="3\msa@28
\mathchardef\therefore="3\msa@29
\mathchardef\because="3\msa@2A
\mathchardef\doteqdot="3\msa@2B

\mathchardef\triangleq="3\msa@2C
\mathchardef\precsim="3\msa@2D
\mathchardef\lesssim="3\msa@2E
\mathchardef\lessapprox="3\msa@2F
\mathchardef\eqslantless="3\msa@30
\mathchardef\eqslantgtr="3\msa@31
\mathchardef\curlyeqprec="3\msa@32
\mathchardef\curlyeqsucc="3\msa@33
\mathchardef\preccurlyeq="3\msa@34
\mathchardef\leqq="3\msa@35
\mathchardef\leqslant="3\msa@36
\mathchardef\lessgtr="3\msa@37
\mathchardef\backprime="0\msa@38
\mathchardef\risingdotseq="3\msa@3A
\mathchardef\fallingdotseq="3\msa@3B
\mathchardef\succcurlyeq="3\msa@3C
\mathchardef\geqq="3\msa@3D
\mathchardef\geqslant="3\msa@3E
\mathchardef\gtrless="3\msa@3F
\mathchardef\sqsubset="3\msa@40
\mathchardef\sqsupset="3\msa@41
\mathchardef\vartriangleright="3\msa@42
\mathchardef\vartriangleleft="3\msa@43
\mathchardef\trianglerighteq="3\msa@44
\mathchardef\trianglelefteq="3\msa@45
\mathchardef\bigstar="0\msa@46
\mathchardef\between="3\msa@47
\mathchardef\blacktriangledown="0\msa@48
\mathchardef\blacktriangleright="3\msa@49
\mathchardef\blacktriangleleft="3\msa@4A
\mathchardef\vartriangle="0\msa@4D
\mathchardef\blacktriangle="0\msa@4E
\mathchardef\triangledown="0\msa@4F
\mathchardef\eqcirc="3\msa@50
\mathchardef\lesseqgtr="3\msa@51
\mathchardef\gtreqless="3\msa@52
\mathchardef\lesseqqgtr="3\msa@53
\mathchardef\gtreqqless="3\msa@54
\mathchardef\Rrightarrow="3\msa@56
\mathchardef\Lleftarrow="3\msa@57
\mathchardef\veebar="2\msa@59
\mathchardef\barwedge="2\msa@5A
\mathchardef\doublebarwedge="2\msa@5B
\mathchardef\angle="0\msa@5C
\mathchardef\measuredangle="0\msa@5D
\mathchardef\sphericalangle="0\msa@5E
\mathchardef\varpropto="3\msa@5F
\mathchardef\smallsmile="3\msa@60
\mathchardef\smallfrown="3\msa@61
\mathchardef\Subset="3\msa@62
\mathchardef\Supset="3\msa@63
\mathchardef\Cup="2\msa@64

\mathchardef\Cap="2\msa@65

\mathchardef\curlywedge="2\msa@66
\mathchardef\curlyvee="2\msa@67
\mathchardef\leftthreetimes="2\msa@68
\mathchardef\rightthreetimes="2\msa@69
\mathchardef\subseteqq="3\msa@6A
\mathchardef\supseteqq="3\msa@6B
\mathchardef\bumpeq="3\msa@6C
\mathchardef\Bumpeq="3\msa@6D
\mathchardef\lll="3\msa@6E

\mathchardef\ggg="3\msa@6F

\mathchardef\circledS="0\msa@73
\mathchardef\pitchfork="3\msa@74
\mathchardef\dotplus="2\msa@75
\mathchardef\backsim="3\msa@76
\mathchardef\backsimeq="3\msa@77
\mathchardef\complement="0\msa@7B
\mathchardef\intercal="2\msa@7C
\mathchardef\circledcirc="2\msa@7D
\mathchardef\circledast="2\msa@7E
\mathchardef\circleddash="2\msa@7F
\def\ulcorner{\delimiter"4\msa@70\msa@70 }
\def\urcorner{\delimiter"5\msa@71\msa@71 }
\def\llcorner{\delimiter"4\msa@78\msa@78 }
\def\lrcorner{\delimiter"5\msa@79\msa@79 }
\def\yen{\mathhexbox\msa@55 }
\def\checkmark{\mathhexbox\msa@58 }
\def\circledR{\mathhexbox\msa@72 }
\def\maltese{\mathhexbox\msa@7A }
\mathchardef\lvertneqq="3\msb@00
\mathchardef\gvertneqq="3\msb@01
\mathchardef\nleq="3\msb@02
\mathchardef\ngeq="3\msb@03
\mathchardef\nless="3\msb@04
\mathchardef\ngtr="3\msb@05
\mathchardef\nprec="3\msb@06
\mathchardef\nsucc="3\msb@07
\mathchardef\lneqq="3\msb@08
\mathchardef\gneqq="3\msb@09
\mathchardef\nleqslant="3\msb@0A
\mathchardef\ngeqslant="3\msb@0B
\mathchardef\lneq="3\msb@0C
\mathchardef\gneq="3\msb@0D
\mathchardef\npreceq="3\msb@0E
\mathchardef\nsucceq="3\msb@0F
\mathchardef\precnsim="3\msb@10
\mathchardef\succnsim="3\msb@11
\mathchardef\lnsim="3\msb@12
\mathchardef\gnsim="3\msb@13
\mathchardef\nleqq="3\msb@14
\mathchardef\ngeqq="3\msb@15
\mathchardef\precneqq="3\msb@16
\mathchardef\succneqq="3\msb@17
\mathchardef\precnapprox="3\msb@18
\mathchardef\succnapprox="3\msb@19
\mathchardef\lnapprox="3\msb@1A
\mathchardef\gnapprox="3\msb@1B
\mathchardef\nsim="3\msb@1C
\mathchardef\ncong="3\msb@1D

\mathchardef\varsubsetneq="3\msb@20
\mathchardef\varsupsetneq="3\msb@21
\mathchardef\nsubseteqq="3\msb@22
\mathchardef\nsupseteqq="3\msb@23
\mathchardef\subsetneqq="3\msb@24
\mathchardef\supsetneqq="3\msb@25
\mathchardef\varsubsetneqq="3\msb@26
\mathchardef\varsupsetneqq="3\msb@27
\mathchardef\subsetneq="3\msb@28
\mathchardef\supsetneq="3\msb@29
\mathchardef\nsubseteq="3\msb@2A
\mathchardef\nsupseteq="3\msb@2B
\mathchardef\nparallel="3\msb@2C
\mathchardef\nmid="3\msb@2D
\mathchardef\nshortmid="3\msb@2E
\mathchardef\nshortparallel="3\msb@2F
\mathchardef\nvdash="3\msb@30
\mathchardef\nVdash="3\msb@31
\mathchardef\nvDash="3\msb@32
\mathchardef\nVDash="3\msb@33
\mathchardef\ntrianglerighteq="3\msb@34
\mathchardef\ntrianglelefteq="3\msb@35
\mathchardef\ntriangleleft="3\msb@36
\mathchardef\ntriangleright="3\msb@37
\mathchardef\nleftarrow="3\msb@38
\mathchardef\nrightarrow="3\msb@39
\mathchardef\nLeftarrow="3\msb@3A
\mathchardef\nRightarrow="3\msb@3B
\mathchardef\nLeftrightarrow="3\msb@3C
\mathchardef\nleftrightarrow="3\msb@3D
\mathchardef\divideontimes="2\msb@3E
\mathchardef\varnothing="0\msb@3F
\mathchardef\nexists="0\msb@40
\mathchardef\mho="0\msb@66
\mathchardef\eth="0\msb@67
\mathchardef\eqsim="3\msb@68
\mathchardef\beth="0\msb@69
\mathchardef\gimel="0\msb@6A
\mathchardef\daleth="0\msb@6B
\mathchardef\lessdot="3\msb@6C
\mathchardef\gtrdot="3\msb@6D
\mathchardef\ltimes="2\msb@6E
\mathchardef\rtimes="2\msb@6F
\mathchardef\shortmid="3\msb@70
\mathchardef\shortparallel="3\msb@71
\mathchardef\smallsetminus="2\msb@72
\mathchardef\thicksim="3\msb@73
\mathchardef\thickapprox="3\msb@74
\mathchardef\approxeq="3\msb@75
\mathchardef\succapprox="3\msb@76
\mathchardef\precapprox="3\msb@77
\mathchardef\curvearrowleft="3\msb@78
\mathchardef\curvearrowright="3\msb@79
\mathchardef\digamma="0\msb@7A
\mathchardef\varkappa="0\msb@7B
\mathchardef\hslash="0\msb@7D
\mathchardef\hbar="0\msb@7E
\mathchardef\backepsilon="3\msb@7F
\def\Bbb{\ifmmode\let\next\Bbb@\else
 \def\next{\errmessage{Use \string\Bbb\space only in math mode}}\fi\next}
\def\Bbb@#1{{\Bbb@@{#1}}}
\def\Bbb@@#1{\fam\msbfam#1}
\font\sacfont=eufm10 scaled 1440
\catcode`\@=12

\def\sla#1{\mkern-1.5mu\raise0.4pt\hbox{$\not$}\mkern1.2mu #1\mkern 0.7mu}
\def\Dbar{\mkern-1.5mu\raise0.4pt\hbox{$\not$}\mkern-.1mu {\rm D}\mkern.1mu}
\def\Abar{\mkern1.mu\raise0.4pt\hbox{$\not$}\mkern-1.3mu A\mkern.1mu}
\else\message{(This uses pseudo 12pts fonts.}
\hoffset=8mm
\voffset=16mm
\input lfont12 

\def\sla#1{\mkern-1.5mu\raise0.5pt\hbox{$\not$}\mkern1.2mu #1\mkern 0.7mu}
\def\Dbar{\mkern-1.5mu\raise0.5pt\hbox{$\not$}\mkern-.1mu {\rm D}\mkern.1mu}
\def\Abar{\mkern1.mu\raise0.5pt\hbox{$\not$}\mkern-1.3mu A\mkern.1mu}
\fi
\def\fileth{\noalign{\hrule}}

\newcount\yearltd\yearltd=\year\advance\yearltd by -1900
\newif\ifdraftmode
\draftmodefalse
\def\draft{\draftmodetrue{\count255=\time\divide\count255 by 60
\xdef\hourmin{\number\count255} 
  \multiply\count255 by-60\advance\count255 by\time
  \xdef\hourmin{\hourmin:\ifnum\count255<10 0\fi\the\count255}}}
 
\newif\iffrancmode
\francmodefalse
\def\e{\mathop{\rm e}\nolimits}

\def\d{{\rm d}}
\def\ud{{\textstyle{1\over 2}}}

\def\del{\partial}

\chardef\sigmat=27

\def\frac#1#2{{\textstyle{#1\over#2}}}

\def\today{\number\day/\number\month/\number\year}
\def\leaderfill{\leaders\hbox to 1em{\hss.\hss}\hfill}
\catcode`\@=11
\def\deqalignno#1{\displ@y\tabskip\centering \halign to
\displaywidth{\hfil$\displaystyle{##}$\tabskip0pt&$\displaystyle{{}##}$
\hfil\tabskip0pt &\quad
\hfil$\displaystyle{##}$\tabskip0pt&$\displaystyle{{}##}$ 
\hfil\tabskip\centering& \llap{$##$}\tabskip0pt \crcr #1 \crcr}}
\def\deqalign#1{\null\,\vcenter{\openup\jot\m@th\ialign{
\strut\hfil$\displaystyle{##}$&$\displaystyle{{}##}$\hfil
&&\quad\strut\hfil$\displaystyle{##}$&$\displaystyle{{}##}$
\hfil\crcr#1\crcr}}\,}
\newread\ch@ckfile
\def\cinput#1{\def\filen@me{#1}
\immediate\openin\ch@ckfile=\filen@me
\ifeof\ch@ckfile\closein\ch@ckfile\message{<< (\filen@me\ N'EXISTE PAS)
>>}\else%
\input\filen@me\closein \ch@ckfile\fi}
\immediate\openin\ch@ckfile=\jobname.def
\ifeof\ch@ckfile\closein\ch@ckfile\message{<< (\jobname.def N'EXISTE PAS)
>>}
\def\DefWarn#1{\ifx\UNd@FiNeD#1\else
\immediate\write16{*** WARNING: the label \string#1 is already defined%
***}\fi}%
\else%
\def\DefWarn#1{}
\input\jobname.def\closein \ch@ckfile\fi
\newcount\nosection
\newcount\nosubsection
\newcount\neqno
\newcount\notenumber
\newcount\nofigure
\newif\ifappmode
\def\table{\jobname.toc}
\def\equation{\jobname.equ}
\def\labeldefs{\jobname.eqd}
\newwrite\equa
\newwrite\tab 
\newwrite\eqdf

\newdimen\hulp
\def\maketitle#1{
\edef\oneliner##1{\centerline{##1}}
\edef\twoliner##1{\vbox{\parindent=0pt\leftskip=0pt plus 1fill\rightskip=0pt
plus 1fill 
                     \parfillskip=0pt\relax##1}} 
\setbox0=\vbox{#1}\hulp=0.5\hsize
                 \ifdim\wd0<\hulp\oneliner{#1}\else
                 \twoliner{#1}\fi}
\def\preprint#1{{\sacfont }\hfill{#1}\vskip 20mm}
\def\title#1\par{\gdef\titlename{#1}
\maketitle{\bf\uppercase\expandafter{\titlename}}
\vskip8truemm
\nosection=0
\neqno=0
\notenumber=0
\nofigure=0
\def\prefix{}
\appmodefalse
\mark{\the\nosection}
\message{#1}
\immediate\openout\equa=\equation
\immediate\openout\eqdf=\labeldefs
}
\def\abstract{\bigskip%
 \begingroup\narrower
\elevenpoint\baselineskip10pt} 
\def\endabstract{\par\endgroup\bigskip
}
\def\section#1\par{\vskip0pt plus.1\vsize\penalty-100\vskip0pt plus-.1
\vsize\bigskip\vskip\parskip
\ifnum\nosection=0\ifappmode\relax\else\writetoc
\fi\fi
\advance\nosection by 1\global\nosubsection=0\global\neqno=0
\vbox{\noindent\bf{\hyperdef\hypernoname{section}{\prefix\the\nosection}%
{\prefix\the\nosection}\ #1}}
\writetoca{{\string\hyperref{}{section}{\prefix\the\nosection}%
{\prefix\the\nosection}} {#1}}
\message{\the\nosection\ #1}
\mark{\the\nosection}\bigskip\noindent
}

\def\appendix#1#2\par{\bigbreak\nosection=0
\appmodetrue
\notenumber=0
\neqno=0
\def\prefix{A}
\mark{\the\nosection}
\message{APPENDICES}
{\leftline{APPENDICES} \hyperdef\hypernoname{appendix}{app}{ 
\leftline{\uppercase\expandafter{#1}}
\leftline{\uppercase\expandafter{#2}}}}
\bigskip\noindent\nonfrenchspacing
\writetoca{\string\hyperref{}{appendix}{app}{Appendices}.\ #1.\ #2}%
}
\def\subsection#1\par {\vskip0pt plus.05\vsize\penalty-100\vskip0pt
plus-.05\vsize\bigskip\vskip\parskip\advance\nosubsection by 1
\vbox{\noindent\it{\hyperdef\hypernoname{subsection}{\prefix\the\nosection.%
\the\nosubsection}{\prefix\the\nosection.\the\nosubsection\ #1}}}%
\smallskip\noindent 
\writetoca{{\string\hyperref{}{subsection}{\prefix\the\nosection.%
\the\nosubsection}{\prefix\the\nosection.\the\nosubsection}} {#1}}
\message{\the\nosection.\the\nosubsection\ #1}
} 
\def\note #1{\advance\notenumber by 1
\footnote{$^{\the\notenumber}$}{\sevenrm #1}} 

\parindent=1em 
\newinsert\margin
\dimen\margin=\maxdimen
\count\margin=0 \skip\margin=0pt
\def\sslbl#1{\DefWarn#1%
\ifdraftmode{\hfill\escapechar-1{\rlap{\hskip-1mm%
\sevenrm\string#1}}}\fi%
\ifnum\nosection=0\xdef#1{}%
\edef\ewrite{\write\eqdf{\string\def\string#1{}}
\write\eqdf{}}\ewrite%
\edef\ewrite{\write\equa{{\string#1}}%
\write\equa{}}\ewrite%
\else%
\ifnum\nosubsection=0%
\xdef#1{\noexpand\hyperref{}{section}{\prefix\the\nosection}{\prefix\the\nosection}}%
\edef\ewrite{\write\eqdf{\string\def\string#1{\prefix%
\the\nosection}}\write\eqdf{}}\ewrite%
\edef\ewrite{\write\equa{{\string#1},\prefix\the\nosection}%
\write\equa{}}\ewrite%
\writedef{#1\leftbracket#1}
\else%
\xdef#1{\noexpand\hyperref{}{subsection}{\prefix\the\nosection.%
\the\nosubsection}{\prefix\the\nosection.\the\nosubsection}}%
\writedef{#1\leftbracket#1}
\edef\ewrite{\write\eqdf{\string\def\string#1{\prefix%
\the\nosection.\the\nosubsection}}\write\eqdf{}}\ewrite%
\edef\ewrite{\write\equa{{\string#1},\prefix\the\nosection%
.\the\nosubsection}\write\equa{}}\ewrite\fi\fi}%
\def\listcontent{\immediate\closeout\tfile\immediate\openin%
\ch@ckfile=\jobname.tab
\ifeof\ch@ckfile\message{no file \jobname.tab, no table of contents this
pass}%
\else\closein\ch@ckfile\centerline{\bf\iffrancmode Table des
mati\`eres \else Contents\fi}\nobreak\medskip%
{\baselineskip=12pt\parskip=0pt\catcode`\@=11\input\jobname.tab
\catcode`\@=12\bigbreak\bigskip}\fi}
\newwrite\tfile \def\writetoca#1{}
\def\writetoc{\immediate\openout\tfile=\jobname.tab
   \def\writetoca##1{{\edef\next{\write\tfile{\noindent ##1
   \string\leaderfill {\string\hyperref{}{page}{\noexpand\number\pageno}%
                       {\noexpand\number\pageno}} \par}}\next}}}

%
\def\nolabels{\def\wrlabeL##1{}\def\eqlabeL##1{}\def\reflabeL##1{}}
\def\writelabels{\def\wrlabeL##1{\leavevmode\vadjust{\rlap{\smash%
{\line{{\escapechar=` \hfill\rlap{\sevenrm\hskip.03in\string##1}}}}}}}%
\def\eqlabeL##1{{\escapechar-1\rlap{\sevenrm\hskip.05in\string##1}}}%
\def\reflabeL##1{\noexpand\llap{\noexpand\sevenrm\string\string\string##1}}}
\nolabels

\global\newcount\refno \global\refno=1
\newwrite\rfile
\def\ref{[\hyperref{}{reference}{\the\refno}{\the\refno}]\nref}
\def\nref#1{\DefWarn#1%
\xdef#1{[\noexpand\hyperref{}{reference}{\the\refno}{\the\refno}]}%
\writedef{#1\leftbracket#1}%
\ifnum\refno=1\immediate\openout\rfile=\jobname.ref\fi
\chardef\wfile=\rfile\immediate\write\rfile{\noexpand\item{[\noexpand\hyperdef%
\noexpand\hypernoname{reference}{\the\refno}{\the\refno}]\ }%
\reflabeL{#1\hskip.31in}\pctsign}\global\advance\refno by1\findarg}
\def\findarg#1#{\begingroup\obeylines\newlinechar=`\^^M\pass@rg}
{\obeylines\gdef\pass@rg#1{\writ@line\relax #1^^M\hbox{}^^M}%
\gdef\writ@line#1^^M{\expandafter\toks0\expandafter{\striprel@x #1}%
\edef\next{\the\toks0}\ifx\next\em@rk\let\next=\endgroup\else\ifx\next\empty%
\else\immediate\write\wfile{\the\toks0}\fi\let\next=\writ@line\fi\next\relax}}
\def\striprel@x#1{} \def\em@rk{\hbox{}}
\def\lref{\begingroup\obeylines\lr@f}
\def\lr@f#1#2{\DefWarn#1\gdef#1{\let#1=\UNd@FiNeD\ref#1{#2}}\endgroup\unskip}

\def\addref#1{\immediate\write\rfile{\noexpand\item{}#1}} 
\def\listrefs{{}\vfill\supereject\immediate\closeout\rfile\writestoppt
\baselineskip=14pt\centerline{{\bf\iffrancmode R\'eferences\else References%
\fi}}
\bigskip{\parindent=20pt%
\frenchspacing\escapechar=` \input \jobname.ref\vfill\eject}\nonfrenchspacing}
\def\startrefs#1{\immediate\openout\rfile=\jobname.ref\refno=#1}
\def\xref{\expandafter\xr@f}\def\xr@f[#1]{#1}
\def\refs#1{\count255=1[\r@fs #1{\hbox{}}]}
\def\r@fs#1{\ifx\UNd@FiNeD#1\message{reflabel \string#1 is undefined.}%
\nref#1{need to supply reference \string#1.}\fi%
\vphantom{\hphantom{#1}}{\let\hyperref=\relax\xdef\next{#1}}%
\ifx\next\em@rk\def\next{}%
\else\ifx\next#1\ifodd\count255\relax\xref#1\count255=0\fi%
\else#1\count255=1\fi\let\next=\r@fs\fi\next}
%
\newwrite\lfile
{\escapechar-1\xdef\pctsign{\string\%}\xdef\leftbracket{\string\{}
\xdef\rightbracket{\string\}}\xdef\numbersign{\string\#}}
\def\writedefs{\immediate\openout\lfile=\jobname.def \def\writedef##1{%
{\let\hyperref=\relax\let\hyperdef=\relax\let\hypernoname=\relax
 \immediate\write\lfile{\string\def\string##1\rightbracket}}}}%
\def\writestop{\def\writestoppt{\immediate\write\lfile{\string\pageno%
\the\pageno\string\startrefs\leftbracket\the\refno\rightbracket%
\string\def\string\secsym\leftbracket\secsym\rightbracket%
\string\secno\the\secno\string\meqno\the\meqno}\immediate\closeout\lfile}}
\def\writestoppt{}\def\writedef#1{}
\writedefs
\def\biblio\par{\vskip0pt plus.1\vsize\penalty-100\vskip0pt plus-.1
\vsize\bigskip\vskip\parskip
\message{Bibliographie}
{\leftline{\bf \hyperdef\hypernoname{biblio}{bib}{Bibliographical Notes}}}
\nobreak\medskip\noindent\frenchspacing
\writetoca{\string\hyperref{}{biblio}{bib}{Bibliographical Notes}}}%

\def\biblionote{\iffrancmode Notes Bibliographiques\else Bibliographical Notes
\fi}
\def\beginbib\par{\vskip0pt plus.1\vsize\penalty-100\vskip0pt plus-.1
\vsize\bigskip\vskip\parskip
\message{Bibliographie}
{\leftline{\bf \hyperdef\hypernoname{biblio}{\the\nosection}%
{\biblionote}}}
\nobreak\medskip\noindent\frenchspacing
\writetoca{\string\hyperref{}{biblio}{\the\nosection}%
{\biblionote}}}%

\def\Exercises{\iffrancmode Exercices\else Exercises
\fi}
\def\exerc\par{\vskip0pt plus.1\vsize\penalty-100\vskip0pt plus-.1
\vsize\bigskip\vskip\parskip
\message{Exercises}
{\leftline{\bf \hyperdef\hypernoname{exercise}{\the\nosection}{\Exercises}}}
\nobreak\medskip\noindent\frenchspacing
\writetoca{\string\hyperref{}{exercise}{\the\nosection}{\Exercises}}
}
\def\eqnn{\global\advance\neqno by 1 \ifinner\relax\else%
\eqno\fi(\prefix\the\nosection.\the\neqno)}
%
\def\eqnd#1{\global\advance\neqno by 1 
{\xdef#1{($\noexpand\hyperref{}{equation}{\prefix\the\nosection.\the\neqno}%
{\prefix\the\nosection.\the\neqno}$)}}
\ifinner\relax\else\eqno\fi(\hyperdef\hypernoname{equation}{\prefix\the%
\nosection.\the\neqno}{\prefix\the\nosection.\the\neqno})
\writedef{#1\leftbracket#1}
\ifdraftmode{\escapechar-1{\rlap{\hskip.2mm\sevenrm\string#1}}}\fi
\edef\ewrite{\write\eqdf{\string\def\string#1{($\prefix\the\nosection.%
\the\neqno$)}}%
\write\eqdf{}}\ewrite%
\edef\ewrite{\write\equa{{\string#1},(\prefix\the\nosection.\the\neqno)
{\noexpand\number\pageno}}\write\equa{}}\ewrite}
%
\def\checkm@de#1#2{\ifmmode{\def\f@rst##1{##1}\hyperdef\hypernoname{equation}%
{#1}{#2}}\else\hyperref{}{equation}{#1}{#2}\fi}
\def\f@rst#1{\c@t#1a\em@ark}\def\c@t#1#2\em@ark{#1}
\def\eqna#1{\global\advance\neqno by1\ifdraftmode{\hfill%
\escapechar-1{\rlap{\sevenrm\string#1}}}\fi%
\xdef #1##1{(\noexpand\relax\noexpand%
\checkm@de{\prefix\the\nosection.\the\neqno\noexpand\f@rst{##1}1}%
{\hbox{$\prefix\the\nosection.\the\neqno##1$}})}
\writedef{#1\numbersign1\leftbracket#1{\numbersign1}}%
} 
%

%
\def\em@rk{\hbox{}} 
\def\xeqn{\expandafter\xe@n}\def\xe@n(#1){#1}
\def\xeqna#1{\expandafter\xe@na#1}\def\xe@na\hbox#1{\xe@nap #1}
\def\xe@nap$(#1)${\hbox{$#1$}}
\def\eqns#1{(\e@ns #1{\hbox{}})}
\def\e@ns#1{\ifx\UNd@FiNeD#1\message{eqnlabel \string#1 is undefined.}%
\xdef#1{(?.?)}\fi{\let\hyperref=\relax\xdef\next{#1}}%
\ifx\next\em@rk\def\next{}%
\else\ifx\next#1\xeqn#1\else\def\n@xt{#1}\ifx\n@xt\next#1\else\xeqna#1\fi
\fi\let\next=\e@ns\fi\next}
\def\figure#1#2{\global\advance\nofigure by 1 \vglue#1%
{\elevenpoint
\setbox1=\hbox{#2}
\ifdim\wd1=0pt\centerline{Fig.\ \the\nofigure\hskip0.5mm}%
\else\def\caption{Fig.\ \the\nofigure\quad#2\hskip0mm}
\setbox0=\hbox{\caption}
\ifdim\wd0>\hsize\noindent Fig.\ \the\nofigure\quad#2\else
                 \centerline{\caption}\fi\fi}\par}
\def\lfigure#1#2{\global\advance\nofigure by
1\vglue#1\leftline{\elevenpoint\hskip10truemm  Fig.\
\the\nofigure\quad #2}} 
\catcode`@=12

\def\draftend{\vfill\supereject%
\immediate\closeout\equa\immediate\closeout\tab
\ifdraftmode
{\bf \titlename},\par ------------ Date \today. -----------\par
\edef\ewrite{\write\eqdf{}}\ewrite%
\catcode`\&=0
\catcode`\\=10
\input \equation
\catcode`\\=0
\catcode`\&=4\fi
\end
}

\francmodefalse
\def\r{{\rm R}}

\def\cite{\refs} 

\newskip\tableskipamount \tableskipamount=8pt plus 3pt minus 3pt
\def\tableskip{\vskip\tableskipamount}

\preprint{SPhT-98/110}

\title{Determination of critical exponents and equation of state by field
theory methods}

\centerline{\caprm J.~ZINN-JUSTIN}
\medskip
{\capit\centerline{ CEA-Saclay, Service de Physique Th\'eorique*, F-91191
Gif-sur-Yvette,}\par
\centerline{ Cedex, FRANCE} \par
\centerline{E-mail: zinn@spht.saclay.cea.fr}}
 
\footnote{}{Talk given at the 6th International Conference ``Path 
 Integrals: From PeV To TeV'', Florence 25--29 Aug.~1998} 
\footnote{}{${}^{*}$Laboratoire de la Direction des
Sciences de la Mati\`ere du 
Commissariat \`a l'Energie Atomique}

\abstract

Path integrals have played a fundamental role in emphasizing the profound
analogies between Quantum Field Theory (QFT), and Classical as well as
Quantum Statistical Physics. Ideas coming from Statistical Physics
have then led to a deeper understanding of Quantum Field Theory and
open the way for a wealth of non-perturbative methods. Conversely QFT
methods are become essential for the description of the  phase transitions and
critical phenomena beyond mean field theory. This is the point
we want to illustrate here. We therefore review the methods,
based on renormalized $\phi^4_3$ 
quantum field theory and renormalization group, which have led to an
accurate determination of critical exponents of the $N$-vector model,
and more recently of the equation of state of the 3D Ising model. The
starting point is the perturbative expansion for RG functions or the effective
potential to the order presently available. Perturbation theory is known to be
divergent and its divergence has been related to instanton contributions. 
This has allowed to characterize the large order behaviour of perturbation
series, an information that can be used to efficiently ``sum" them. Practical
summation methods based on Borel transformation and 
conformal mapping have been developed, leading to the most accurate results
available probing field theory in a non-perturbative regime. We illustrate
the methods with a short discussion of the scaling equation of
state of the 3D Ising model \ref\rGuiZJ{R.~Guida and J.~Zinn-Justin, {\it
Nucl. Phys.} B489 [FS] (1997) 626; hep-th/9610223.}. Compared to
exponents its determination involves a few additional (non-trivial) technical
steps, like the use of the parametric representation, and the order
dependent mapping method. A general reference on the subject is \par
J. Zinn-Justin, 1989, {\it Quantum Field Theory and Critical Phenomena}, in
particular chap.~28 of third ed., Clarendon Press (Oxford 1989, third ed.
1996).\par 
\endabstract

\vfill\eject
\section Introduction

Second order phase transitions are continuous phase transitions
where the correlation length diverges. Renormalization group (RG) arguments
\ref\rKWW{An early review is K.G. Wilson and J. Kogut, {\it Phys. Rep.}
12C (1974) 75.}, as well as an analysis, near
dimension four, of the most divergent terms appearing in the expansion around
mean field theory \ref\rBLGZJ{E. Br\'ezin, J.C. Le Guillou and J. Zinn-Justin,
in {\it Phase Transitions and Critical Phenomena} vol. 6, C. Domb and M.S.
Green eds. (Academic Press, London 1976).}, indicate that such
transitions present universal features, i.e.~features independent to a
large extend from the details of the microscopic dynamics. Moreover
all universal quantities can be calculated from renormalizable or
super-renormalizable quantum field theories. For an important class of
physical systems and models (with short range interactions) one is led to a
$\phi^4$-like euclidean field theory with $O(N)$ symmetry. Among those let us
mention statistical properties of polymers, liquid--vapour and binary mixtures
transitions, superfluid Helium, ferromagnets...  
We explain here how critical exponents and other
universal quantities have been calculated with field theory techniques. 
To simplify notation we concentrate on the universality class
of the Ising model (models with ${\Bbb Z}_2$ symmetry).
\medskip
{\it The effective quantum field theory.} The relevant field theory
action  $ {\cal H} \left(\phi \right) $ is
$$ {\cal H} \left(\phi \right)= \int \d ^{d}x \left\lbrace\ud
\left[ \nabla \phi (x) \right]^{2}+\ud r\phi
^{2} (x)+\frac{1}{4!}g_0\Lambda^{4-d}\phi^{4} 
(x) \right\rbrace , \eqnn $$
where $r$ plays the role of the temperature. The mass parameter $\Lambda$
corresponds to the inverse microscopic scale and also appears as a
cut-off in the Feynman diagrams of the perturbative expansion.
\par
For some value $r_c$ the correlation length diverges (the physical mass
vanishes), and the critical domain corresponds to $|t=r-r_c|\ll
\Lambda^2$.  
The study of the critical domain reduces to the study of the large
cut-off behaviour, i.e.~to renormalization theory and the corresponding
renormalization group. However one essential feature of the action
distinguishes it from field theory in the form it was traditionally 
presented in particle physics: the dependence on $\Lambda$
of the coefficient of $\phi^4$ is given {\it a
priori}. In particular in the dimensions of interest, $d<4$, the ``bare"
coupling constant diverges, though the field theory, being
super-renormalizable, requires only a mass renormalization.
\par
To circumvent the problem of the large coupling constant the famous
Wilson--Fisher  $\varepsilon$-expansion \ref\rWilFish{K.G. Wilson and
M.E. Fisher, {\it Phys. Rev.  Lett.} 28 (1972) 240.} has been invented. 
The dimension $d$ is considered as a continuous variable. Setting
$d=4-\varepsilon$ one expands both in $g_0$ and $\varepsilon$. Divergences
then behave like in four dimensions, they are only logarithmic and can be
dealt with. Moreover it is possible to study directly the massless
(or critical) theory. \par
Later it has been proposed by Parisi \ref\rParisi{G. Parisi, {\it Carg\`ese
Lectures 1973}, published in {\it J. Stat. Phys.} 23 (1980) 49.}, to
work at fixed dimension $d<4$, in the massive theory (the massless
theory is IR divergent). One motivation for trying such an approach
 is of practical nature: it is easier to calculate
Feynman diagrams in dimension three than in generic dimensions, and thus more
perturbative orders are available. \par  
The large $\Lambda$ limit then is taken first at
$u_0=g_0\Lambda^{4-d}$ fixed. 
This implies that one first tunes the initial parameters of the
model to remain artificially close to the unstable $u_0=0$ gaussian fixed
point: When the correlation length $\xi$ increases
near $T_c$ one decreases the dimensionless bare quantity $g_0$ as 
$g_0^{1/(4-d)} \propto 1/\xi\Lambda $. Finally one takes the infinite
$u_0$ limit. One then is confronted with a serious technical problem:
perturbation theory is finite but one is interested in the infinite
coupling limit.\par
One thus introduces a field renormalization, $\phi=Z^{1/2} \phi_\r$, 
and a renormalized dimensionless coupling constant $g$ as in four
dimensions. They are implicitly defined by the renormalization
conditions for the 
$\phi_\r$ 1PI correlation functions:   
$$\Gamma^{(2)}_{\rm R} (p;m ,g)  = m^2 +p^2 + O \left(p^4
\right), \quad \Gamma^{(4)}_{\rm R} \left(p_i=0;m ,g \right)  =
m^{4-d} g\,. \eqnd \egrzerom $$ 
The role of renormalizations, however, is here different. When the
corresponding Callan--Symanzik $\beta$-function  has an IR stable zero,
$\beta(g^*)=0$ with $\omega\equiv\beta'(g^*)>0$,  
then the new coupling $g$ has a finite limit $g=g^*$ when the
initial coupling constant $u_0$ becomes large. 
Thus the renormalized coupling $g$ is a more suitable expansion
parameter than $u_0$. \par
Note that the mass parameter $m$, which is proportional to the 
physical mass, or inverse correlation length, of the high temperature phase,
behaves for $t=r-r_c\to 0_+$ as $m\propto t^\nu$, where $\nu$ is the
correlation length exponent. 
\par
In contrast with the $\varepsilon$-expansion however, at fixed dimension three
or two one has no small  
parameter. Therefore accurate determinations of $g^*$ and all other
physical quantities depends on the analytic properties of the series,
in addition to the number of 
terms available. A semi-classical analysis, based on instanton calculus,
unfortunately indicates that perturbation theory in $\phi^4$ field theory is 
divergent.  Therefore to extract any information from perturbation theory a
summation method is required. 
\section Critical exponents. Borel transformation and mapping

The most studied quantities are critical exponents, because they are easier to
calculate. They have been extensively used to compare RG predictions
with other results (experiments, high or low temperature
series expansion, Monte-Carlo simulations). 
The first accurate values of the exponents of the $O(N)$ symmetric
$N$-vector model have been reported in ref.~\ref\rLGZJ{J.C. Le Guillou and J.
Zinn-Justin, {\it Phys. Rev. Lett.} 39 (1977) 95; {\it Phys. Rev.} B21 (1980)
3976.} using six-loop series for RG functions \ref\rNMB{B.G.
Nickel, D.I.Meiron, G.B. Baker, Univ. of Guelph Report 1977.}. 
Perturbative series have been summed using Borel transformation and conformal
mapping. The same ideas have later been applied to the
$\varepsilon$-expansion when five loop series have become available, and
recently to the equation of state. With time the method has been refined
and the efficiency improved by various tricks but the basic principles have
not changed. 
\medskip
{\it Borel transformation and conformal mapping.}
Let $R(g)$ be a quantity given by a perturbation series
$$R(g)=\sum R_k g^k. \eqnn $$
Large order behaviour analysis (instantons) \ref\rLGZJed{
{\it Large Order Behaviour of Perturbation Theory}, {\it Current Physics}
vol.~7, J.C. Le Guillou and J. Zinn-Justin eds., (North-Holland, Amsterdam
1990).} teaches us that, in the $\phi^4$ field theory, $R_k$  behaves like
$ k^s (-a)^k k!$ for $k$ large.
The value of $a>0$ has been determined numerically. One thus introduces
$B(g)$, the Borel transform of $R(g)$, which is defined by
$$B(g)=\sum (R_k /k!)\, g^k\,. \eqnd\eBorseries $$
The function $B(g)$ is analytic at least in a circle with the
singularity closest to the origin located at $z=-1/a$. 
Unlike $R(g)$, $B(g)$ is determined by its series expansion. 
In the sense of formal series, $R(g)$ can be recovered from
$$R(g)=\int_0^\infty \e^{-t}B(gt)\d t\,,\eqnd\eBorLer $$
However, for relation \eBorLer~to make sense as a relation between
functions, and not only between formal series, one must know $B(g)$
on the whole real positive axis. 
This implies that $B(g)$ must be analytic near the axis, a result
proven rigorously \ref\rBorsom{J.P. Eckmann, J.
Magnen and R. S\'en\'eor, {\it Commun. Math. Phys.} 39 (1975) 251;
J.S. Feldman and K. Osterwalder, {\it Ann. Phys. (NY)} 97 (1976) 80.}.
Moreover it is necessary to 
continue analytically the function from the circle to the real positive axis. 
Consideration of general instanton contributions suggests that
$B(g)$ actually is analytic in the cut-plane. Therefore 
the analytic continuation can be obtained from a conformal map of the 
cut-plane onto a circle:
$$z\mapsto u(z)=az/( \sqrt{1+za}+ 1)^2 .\eqnn $$
The function $R(g)$ is then given by the new, hopefully convergent, expansion
$$ R (g)= \sum B_{k}
\int^{\infty}_{0}\e^{-t} \left[ u(gt)\right]^{k} \d t\,.\eqnn $$
\medskip
{\it  Exponents.}
The values of critical exponents obtained from field theory have remained
after about twenty years among the most accurate determinations. Only recently
have consistent, but significantly more accurate, experimental results  been
reported in low gravity superfluid experiments \ref\rLipa{ J.A. Lipa, D.R.
Swanson, J. Nissen, T.C.P. Chui and U.E. Israelson, {\it Phys. Rev. Lett.} 76
(1996) 944.}. Also various numerical simulations \ref\rRaGu{C.F. Baillie, R.
Gupta, K.A. Hawick and G.S. Pawley, {\it Phys. Rev.} B45 (1992) 10438,
and references therein;
B. Li, N. Madras and A.D. Sokal, {\it J. Stat. Phys.} 80 (1995) 661;
S. Caracciolo, M.S. Causo and A. Pelissetto, cond-mat/9703250.} 
and high temperature expansions on the lattice have claimed 
similar accuracies.\par 
The values of critical exponents have recently been updated 
\ref\rGuiZJii{R.~Guida and J.~Zinn-Justin, {\it   
J. Phys.} A31 (1998) 8103, cond-mat/9803240.} because seven-loop 
terms have been obtained for two of the
three RG functions. Some results are displayed above. The main
improvements concern the exponent $\eta$ which was poorly
determined, and the lower value of $\gamma$ for $N=0$ (polymers). 
 \topinsert
$$ \vbox{\elevenpoint\offinterlineskip\tabskip=0pt\halign to \hsize
{& \vrule#\tabskip=0em plus1em & \strut\ # \
& \vrule#& \strut # 
& \vrule#& \strut # 
& \vrule#& \strut # 
& \vrule#& \strut # 
&\vrule#\tabskip=0pt\cr
\noalign{\centerline{\it Critical exponents of the
$O(N)$ models from $d=3$ expansion \rGuiZJii.} \tableskip}
\fileth
height2.0pt& \omit&& \omit&& \omit&&\omit&& \omit&\cr
&$ \hfill N \hfill$&&$ \hfill 0
\hfill$&&$ \hfill 1 \hfill$&&$ \hfill 2
\hfill$&&$ \hfill 3 \hfill$&\cr
height2.0pt& \omit&& \omit&& \omit&& \omit&& \omit&\cr
\fileth
height2.0pt& \omit&& \omit&& \omit&& \omit&& \omit&\cr
height2.0pt& \omit&& \omit&& \omit&& \omit&& \omit&\cr
&$ \hfill g^* \hfill$
&&$ \hfill  26.63\pm 0.11  \hfill$
&&$ \hfill 23.64\pm0.07 \hfill$
&&$ \hfill 21.16\pm 0.05 \hfill$
&&$\hfill 19.06\pm0.05 \hfill$&\cr 
height2.0pt& \omit&& \omit&& \omit&& \omit&& \omit&\cr
&$ \hfill \gamma \hfill$
&&$ \hfill 1.1596\pm0.0020 \hfill$
&&$ \hfill 1.2396\pm 0.0013\hfill$
&&$ \hfill 1.3169\pm0.0020  \hfill$
&&$\hfill 1.3895\pm0.0050 \hfill$&\cr
height2.0pt& \omit&& \omit&& \omit&& \omit&& \omit&\cr
&$ \hfill \nu \hfill$
&&$ \hfill 0.5882\pm 0.0011 \hfill$
&&$ \hfill 0.6304\pm 0.0013\hfill$
&&$ \hfill 0.6703\pm 0.0015  \hfill$
&&$\hfill 0.7073\pm 0.0035 \hfill$&\cr
height2.0pt& \omit&& \omit&& \omit&& \omit&&\omit&\cr
&$ \hfill \eta \hfill$
&&$ \hfill 0.0284\pm0.0025\hfill$
&&$ \hfill 0.0335\pm0.0025\hfill$
&&$ \hfill 0.0354\pm0.0025  \hfill$
&&$\hfill 0.0355\pm0.0025 \hfill$&\cr
height2.0pt& \omit&& \omit&& \omit&& \omit&& \omit&\cr
&$ \hfill \beta \hfill$
&&$ \hfill 0.3024\pm0.0008\hfill$
&&$ \hfill 0.3258\pm0.0014\hfill$
&&$ \hfill 0.3470\pm0.0016  \hfill$
&&$\hfill 0.3662\pm0.0025 \hfill$&\cr
height2.0pt& \omit&& \omit&& \omit&& \omit&& \omit&\cr
&$ \hfill \omega \hfill$&
&$ \hfill 0.812\pm0.016  \hfill$&
&$ \hfill 0.799\pm0.011\hfill$&
&$ \hfill 0.789\pm 0.011 \hfill$&
&$\hfill 0.782\pm0.0013 \hfill$&\cr
height2.0pt& \omit&& \omit&& \omit&& \omit&& \omit&\cr
\fileth }}$$
\endinsert
\section 3D Ising model: the scaling equation of state 

Let us first recall a few properties of the equation of the state in
the critical domain, in the specific case $N=1$ (Ising-like
systems), at $d=3$.\sslbl\ssscaleqst 
\par 
The equation of state is the relation between 
magnetic field $H$, magnetization $M=\left<\phi\right>$ (the ``bare"
field expectation value) and the temperature which is represented by the
parameter $t= r-r_c\propto T-T_c$. It is related to
the free energy per unit volume, in field theory
language the generating functional $\Gamma(\phi)$ of 1PI correlation
functions restricted to constant fields, i.e~the effective potential $V$,
$V(M)=\Gamma(M)/{\rm vol.}$, by $H=\del V /\del M$.
In the critical domain the equation of state has Widom's scaling form
\par
$$H(M,t)=M^\delta f(t/M^{1/\beta}), \eqnd\ehscal $$ 
a form initially conjectured and which renormalization group has justified. 
\par
One property of the function $H(M,t)$ which plays an essential 
role in the analysis is {\it Griffith's analyticity}: 
it is regular at $t=0$ for $M>0$ fixed, and simultaneously it is regular at
$M=0$  for $t>0$ fixed. 
\par
In the framework of the $\varepsilon$-expansion, the function $f(x)$ 
has been determined up to order $\varepsilon^2$ for the general $O(N)$ model,
\ref\rBWW{G.M. Avdeeva and A.A. Migdal, {\it JETP Lett.} 16 (1972) 178;
E. Br\'ezin, D.J. Wallace and K.G. Wilson, {\it Phys. Rev. Lett.} 29 (1972)
591; {\it Phys. Rev.} B7 (1973) 232.} and order $\varepsilon^3$ for $N=1$
\ref\rWZAN{D.J. Wallace and R.P.K. Zia, {\it J.
Phys. C: Solid State Phys.} 7 (1974) 3480.}. \par
The calculations presented here are performed within the framework of the 
$\phi^4_3$ massive field theory renormalized at zero momentum
(eq.~\egrzerom). Five loop series for the effective potential have
been  reported  
\ref\rBBMN{C. Bagnuls, C. Bervillier, D.I. Meiron and 
B.G. Nickel, {\it Phys. Rev.} B35 (1987) 3585;
F.J. Halfkann and V. Dohm, {\it Z. Phys.} B89 (1992) 79.}.
The conditions \egrzerom~imply that the effective potential $V$
expressed in terms of the expectation value of the renormalized field
$\varphi=\left<\phi_\r \right>$, has a small $\varphi$ expansion of the form
$$V(\varphi)-V(0)=\ud m^2
\varphi^2+\frac{1}{4!}m g \varphi^4 +O\left(\varphi^6\right) =
(m^3/g){\cal V}(z,g) , \eqnn $$
where $z$ is a dimensionless variable $z=\varphi \sqrt{g/m}$. For
$g=g^*$
$$ z \propto M/\sqrt{mZ}\propto
M/m^{(1+\eta)/2}\propto Mt^{-\beta}.\eqnd\ezscalvar $$ 
The equation of state is related to the derivative $F$ of the reduced 
effective potential $\cal V$ with respect to $z$ 
$$H\propto t^{\beta\delta}F(z) ,\quad F(z)\equiv F(z,g^*)={\partial {\cal
V}(z,g^*)\over \partial z}\,.\eqnn $$ 
\medskip
{\it The problem of the low temperature phase.}
To determine the equation of state in the whole physical range a new
problem arises. In this framework it is
difficult to calculate physical quantities in the ordered phase because
the theory is parametrized in terms of the disordered phase correlation length
$\xi=m^{-1}\propto t^{-\nu}$ which is singular at $T_c$
(as well as all correlation functions normalized as in~\egrzerom{}). 
In the limit $m\rightarrow 0$, at $\varphi$ fixed, $z\to \infty$ as
seen in eq.~\ezscalvar. 
In this limit from eq.~\ehscal~ one finds
$$H(M,t=0)\propto M^{\delta}\ \Rightarrow\ F(z)\propto
z^\delta.\eqnd\ettozero $$
The perturbative expansion of the scaling equation of state leads
to an expression only adequate for the
description of the disordered phase.
\par
In the case of the $\varepsilon$-expansion the scaling relations (and thus the
limiting behaviour~\ettozero) are exactly satisfied order by order.
Moreover the change to the  variable $x\propto z^{-1/\beta}$ (more
appropriate for the regime $t\rightarrow 0$) gives an expression for 
$f(x)\propto F(x^{-\beta}) x^{\beta\delta}$ that is explicitly regular in
$x=0$ (Griffith's analyticity). Still, even there a numerical problem arises
when $\varepsilon=1$ is set.
\par
In the case of fixed dimension perturbation theory, instead, because
IR scaling is obtained only for $g=g^*$ and not for generic values of
$g$, scaling properties are not satisfied at any finite order in $g$.  \par
Several approaches can be used to deal with the problem of continuation to the
ordered phase. A rather powerful method, motivated by the results obtained
within the $\varepsilon$-expansion scheme, is based on the parametric
representation. 
\section Parametric representation of the equation of state and ODM

Both the scaling and regularity properties of the equation of state
can be more easily expressed by parametrizing it in terms of two new
variables $R$ and $\theta$, setting:  
$$M = m_0 R^{\beta}\theta\, , \quad t = R
\left(1-\theta^2 \right),  \quad H  =h_0 R^{\beta\delta}h(\theta)\, ,
 \eqnd\ezmaptheta $$
where $h_0, m_0$ are normalization constants.
Then the function $h(\theta)$   
is an odd function of $\theta$ which from Griffith's analyticity is regular
near $\theta =1$, which is $z$ large, and near $\theta=0$ which is $z$ small. 
It vanishes for $\theta=\theta_0$ which corresponds to the coexistence curve
$H=0,T<T_c$. In terms of the scaling variable $z$ used previously one
finds 
$$z=\rho \theta/(1-\theta^2)^\beta,\quad
h(\theta)=(1-\theta^2)^{\beta\delta} F(z(\theta)), \eqnd\eqstODM $$
where $\rho$ is an arbitrary parameter.  
Note if we know a few terms of the expansion of $F(z)$
in powers of $z$ we know the same number of terms in the expansion of
$h(\theta)$ in powers of $\theta$. But because $h$ is a more regular
function, the latter expansion has a much larger domain of validity. 
\par
From the parametric representation of the equation of state it is then
possible to derive a representation for the singular part of the free energy
per unit volume as well as various universal ratios of amplitudes.
\nref\rNA{J.F. Nicoll and P.C. Albright, {\it Phys. Rev.} B31
(1985) 4576.} 
\medskip
{\it Order dependent mapping (ODM).} In the framework outlined before,
the approximate $h(\theta)$ that one obtains by summing perturbation theory at 
fixed dimension, is still not regular. The terms singular at
$\theta=1$, generated by the mapping \eqstODM, do not cancel exactly due
to summation errors. The last step thus is to Taylor expand
the approximate expression for $h(\theta)$ around $\theta=0$ and to 
truncate the expansion, enforcing in this way regularity. A question
then arises, to which order in $\theta$ should one expand?
Since the coefficients of the $\theta$ expansion are in one to one
correspondence with the coefficients of the small $z$ expansion of the
function $F(z)$, the maximal power of $\theta$ in $h(\theta )$, 
should be equal to the maximal power of $z$ whose coefficient can be
determined with reasonable accuracy. As noted before, although the
small $z$ expansion of $F(z)$ at each finite loop order in $g$
contains an infinite number of terms, the evaluation of the
coefficients of the higher powers of $z$ is increasingly difficult.\par
Therefore one has to ensure the fastest possible
convergence of the small $\theta$ expansion. For this purpose one uses the
freedom in the choice of the parameter $\rho$ in eq.~\eqstODM:
one determines $\rho$ by minimizing the last term in the truncated small
$\theta$ expansion, thus increasing the importance of small powers of $\theta$
which are more accurately calculated. This is nothing but the application to
this particular example of the series summation method based on ODM
\ref\rOMD{R. 
Seznec and J. Zinn-Justin, {\it J. Math. Phys.} 20 (1979) 1398;
J.C. Le Guillou and J. Zinn-Justin, {\it Ann. Phys. (NY)} 147 (1983) 57;
R. Guida, K. Konishi and H. Suzuki, {\it Ann. Phys. (NY)} 241 (1995) 152; 249
(1996) 109.}.
\section Numerical results

One first determines the first coefficients $F_{2l+1}$ of the small
$z$ (small field) expansion of the function $F(z)$ as accurately as
possible, using the same method as for exponents, i.e.~Borel--Leroy
transformation and conformal mapping. One finds
$$ F_5= 0.01711\pm 0.00007 \,,\quad F_7 \times 10^4=
 4.9\pm 0.5 \,,\quad F_9\times 10^5= -7\pm5 \,.$$
One then determines by the ODM method the  parameter $\rho$ and the function
$h(\theta)$ of the parametric representation, as explained before. One
obtains successive approximations in the form of polynomials of
increasing degree. At leading order one obtains a polynomial of degree five.
It is not possible to go beyond $h_9(\rho)$ because already $F_9$ is  
poorly determined. Note that one has here a simple test of the
relevance of the ODM method. Indeed, once $h(\theta)$ is determined,
assuming the values of the critical exponents $\gamma$ and $\beta$ one can
recover a function $F(z)$ which has an expansion to all orders in $z$. As a
result one obtains a prediction for the coefficients $F_{2l+1}$ which have not
yet been taken into account to determine $h(\theta)$. The relative difference
between the predicted values and the ones directly calculated gives an idea
about the accuracy of the ODM method. 
The simplest representation of the equation of state, consistent
with all data,  is given by
$$h(\theta)=\theta-0.76201(36)\;\theta^3+8.04(11)\times 10^{-3}\;\theta^5,
\eqnd\emainres $$
(errors on the last digits in parentheses) that is obtained 
from $\rho^2=2.8667\ $.
This expression of $h(\theta)$ has a zero at 
$\theta_0=1.154 $, which corresponds to the coexistence curve.
The coefficient of $\theta^7$ in eq.~\emainres~is smaller than $10^{-3}$.
Note that for the largest value of $\theta^2$ which corresponds to 
$\theta_0^2$, the $\theta^5$ term is still a small correction. 
%
%
\medskip
{\it Concluding remarks.}
Within the framework of renormalized quantum field theory and
renormalization group, the presently available series allow,  after
proper summation, to determine  
accurately critical exponents for the $N$-vector model and the complete scaling
equation of state for 3D Ising-like ($N=1$) systems. In the latter example
additional technical tools, beyond Borel summation methods, are required in
which the parametric representation plays a central role. From the equation of
state new estimates of some amplitude ratios have been deduced which seem 
reasonably consistent with all other available data. 
\medskip
{\bf Acknowledgments.} Hospitality at the Center for Theoretical
Physics, MIT, where this
contribution was written, is gratefully acknowledged
\listrefs
\draftend
\end